\documentclass[3p,times,procedia]{elsarticle}
\usepackage{nupha_ecrc}

\usepackage{wrapfig}
\usepackage{amsmath}
\usepackage{amssymb}
\usepackage{graphicx}
\usepackage{bm}
\usepackage{epsfig}
\usepackage{amsfonts}
\usepackage{color}
\usepackage{mathtools}

\volume{00}

\firstpage{1}

\journalname{Nuclear Physics A}

\runauth{}

\jid{nupha}

\jnltitlelogo{Nuclear Physics A}

\usepackage{amssymb}

\usepackage[figuresright]{rotating}

\begin{document}

\begin{frontmatter}



\dochead{}

\title{In-medium quarkonium properties from a lattice QCD based effective field theory}


\author[a]{Seyong Kim}
\author[b]{Peter Petreczky}
\author[c]{Alexander Rothkopf\corref{cor1}$^{*,}$}
\address[a]{Department of Physics, Sejong University, Seoul 143-747, Korea}
\address[b]{Physics Department, Brookhaven National Laboratory, Upton, NY11973, USA}
\address[c]{Institute for Theoretical Physics, Heidelberg University,
        Philosophenweg 16, 69120 Heidelberg, Germany \vspace{-0.4cm}}
\cortext[cor1]{\vspace{-0.5cm} Speaker}

\begin{abstract}
In order to understand the experimental data on heavy quarkonium production in heavy ion collisions at RHIC and LHC it is necessary (though not sufficient) to pinpoint the properties of heavy $Q\bar{Q}$ bound states in the deconfined quark-gluon plasma, including their dissolution. Here we present recent results on the temperature dependence of bottomonium and charmonium correlators, as well as their spectral functions in a lattice QCD based effective field theory called NRQCD, surveying temperatures close to the crossover transition $140 {\rm MeV} < T< 249 {\rm MeV}$. The spectra are reconstructed based on a novel Bayesian prescription, whose systematic uncertainties are assessed. We present indications for sequential melting of different quarkonium species with respect to their vacuum binding energies and give estimates on the survival of S-wave and P-wave ground states.\end{abstract}


\end{frontmatter}



\vspace{0.5cm}

The bound states of a heavy quark and anti-quark, so called heavy quarkonium ($c\bar{c}$ Charmonium, $b\bar{b}$ Bottomonium) are particularly well suited probes to shed light on the properties of the quark gluon plasma (QGP) created in relativistic heavy-ion collisions at RHIC and LHC \cite{Mocsy:2013syh}. On the one hand their di-lepton decay channels allow measurements of yields and spectra with high precision and on the other hand the fact that the heavy quark rest mass ($m_c=1.275(25)$GeV, $m_b=4.66(3)$GeV) lies well above the scales present in a typical QGP created in such a collision $(T\sim\Lambda_{\rm QCD}\sim 200{\rm MeV})$, allows for powerful simplifying effective field theory (EFT) methods.

Recent di-lepton measurements by CMS \cite{Chatrchyan:2011pe} and ALICE \cite{Abelev:2012rv} during run1 at the LHC have revealed two distinct faces of heavy quarkonium, depending on the flavors involved. Comparison of $b\bar{b}$ di-muon spectra in $p+p$ and $Pb+Pb$ collisions indicates that in the presence of a hot medium, less deeply bound states are affected more strongly compared to the ground state. Such a sequential suppression pattern is reminiscent of theory predictions \cite{Matsui:1986dk,Karsch:2005nk} based on a picture of $b\bar{b}$ pairs created during the partonic stages of the collision quickly forming a bound state and then traversing the medium as a well defined colorless probe. Secondly, the Charmonium vector channel ground state $J/\psi$, as measured by ALICE at mid rapidity, shows a replenishment of yields with increasing number of collision participants. Interpreted in the context of the statistical model of hadronization \cite{BraunMunzinger:2000px}, such behavior hints at the importance of uncorrelated pairs of $c$ and $\bar{c}$ recombining into a $J/\psi$ at the freezeout boundary. A first principles understanding of either two scenarios has not yet been fully achieved and constitutes one central focus for theory (see e.g. \cite{Aarts:2014cda,Kim:2014iga}).

For a realistic modeling of quarkonium production in heavy ion collisions based on the idea of sequential suppression \cite{Matsui:1986dk} and statistical regeneration \cite{BraunMunzinger:2000px} detailed information on in-medium $Q\bar{Q}$
properties is needed. This can be provided by first principles lattice calculation of quarkonium correlators and  spectral functions.

Computing $Q\bar{Q}$ correlation functions in the vicinity of the transition is challenging, since the light medium d.o.f. are strongly correlated. Lattice QCD simulations in imaginary time are a proven tool to carry out reliable computations in this temperature range from first principles. As they operate on a discretized spacetime grid, the inclusion of heavy quarks at first seems daunting, since to resolve the energy scale related to the heavy quark mass, $m_Q\gg T$, a much finer discretization is necessary than that for the physics of the light QGP partons itself. This separation of scales can however be turned into an advantage by going over to an effective description of heavy quarks in terms of non-relativistic spinor fields that propagate in the background of a fully relativistic medium, simulated in conventional lattice QCD. This approach, called lattice NRQCD \cite{Lepage:1992tx} is well established at $T=0$ and in principle is applicable also in a medium. Note that it does not involve modeling, as it rests on a systematic expansion of the QCD Lagrangian in powers of the heavy quark velocity $v=\frac{\hat{p}_{\rm Lat}}{m_Q a}$ with $a$ the lattice spacing. 

The only required input is experimental data to set the overall scale, beyond which ab-initio predictions are possible, as was demonstrated by the successful prediction \cite{Dowdall:2011wh} of the $\eta_b(2S)$ vacuum mass before its experimental discovery. Our study benefits from {\it recent progress} in this field, making it possible to simulate the QGP close to the physical point. The lattices used in the following, generated by the HotQCD collaboration \cite{Bazavov:2011nk}, contain dynamical u,d and s quarks with physical mass ratio, feature an $m_\pi=161$MeV, as well as a $T_{\rm pc}=159$MeV, close to the continuum limit value of $T_{\rm pc}^{\rm cont.}=154\pm9$MeV \cite{Bazavov:2011nk}. For the successful application of NRQCD the size of $1/(m_Q a)$ is important, which on these lattices $T\in[140-249]$MeV takes the values $m_b a\in[2.759-1.559]$ and $m_c a\in[0.757-0.427]$.

In practice one computes the in-medium propagator of a single heavy quark $G(\tau,\mathbf{x})$ in the background of the medium as an initial value problem in Euclidean time. By combining the propagators of a quark and antiquark and projecting to the quantum numbers of the state of interest by inserting appropriate combinations of Pauli matrices ($\vec{\mathbf{s}}$) and derivative operators ($\vec{\mathbf{L}}$) a heavy $Q\bar{Q}$ correlator $D(\tau)$ can be constructed. Intuitively it can be regarded as the unequal (imaginary)time correlation function of the $Q\bar{Q}$ wavefunction. Comparing the ratio of these correlators at $T=0$ and $T>0$ provides an estimate of the overall size of the in-medium effects. The modification a single state receives, however, can only be elucidated by extracting from the correlator the corresponding spectral function $\rho(\omega)$. In it bound states appear as Breit-Wigner shaped peaks with position $m_{q\bar{q}}$ and a width that corresponds to the probability of the state to transition to a neighboring bound state or to dissolve into the continuum.

To obtain spectral functions $\rho(\omega)$ from NRQCD correlators $D(\tau)$ one faces an ill-posed problem, where a Laplace transform $D(\tau)=\int\,d\omega\, {\rm exp}[-\omega\tau] \, \rho(\omega)$ has to be inverted. The Euclidean correlator is known only at a discrete number of times $N_\tau$ and with a finite error. To resolve possible bound state features the spectrum needs to be well resolved $N_\omega \gg N_\tau$ and hence the discretized problem $D(\tau_i)=\sum_{l=1}^{N_\omega} \,\Delta\omega_l \, K_{il}\, \rho_l$ cannot be solved by a simple $\chi^2$ fit. It is possible to meaningfully estimate the $\rho_l$ by systematically incorporating additional, so called prior information. The framework of Bayesian inference provides the mathematical tools to do so. A popular implementation of this strategy to regularize a naive likelihood fit based on prior information is the Maximum Entropy Method \cite{Asakawa:2000tr}, which however is known to be susceptible to artificial smoothening if an implementation with a singular value decomposition is used.  The {\it recent construction} of a novel Bayesian approach \cite{Burnier:2013nla} allows to circumvent some of these impediments of MEM and has been shown to be able to produce up to an order of magnitude better resolved spectra from the same correlators.

Since the underlying problem is ill-defined any two Bayesian approaches using a finite number of datapoints $N_\tau$ and finite error $\Delta D/D$ will produce different outcomes. Only in the limit  $N_\tau\to\infty$ and $\Delta D/D\to0$ will they agree. A benefit of the novel Bayesian approach due to the absence of flat directions is that larger $N_\tau$ does not lead to slow convergence and thus the approach to these limits can be explored in practice.

\begin{wrapfigure}{r}{0.73\textwidth}
  \begin{center}\vspace{-1.cm}
   \includegraphics[scale=0.65]{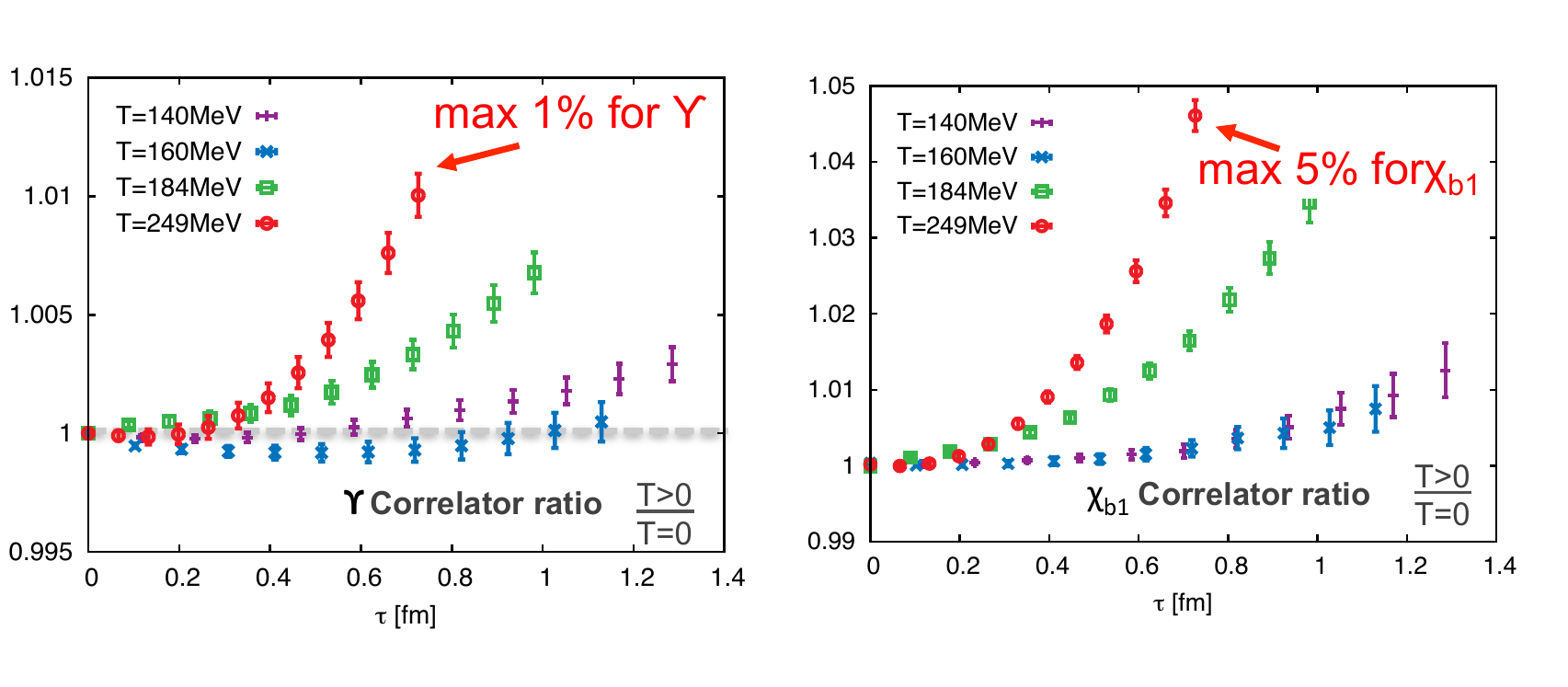}\vspace{-0.6cm}
  \includegraphics[scale=0.675]{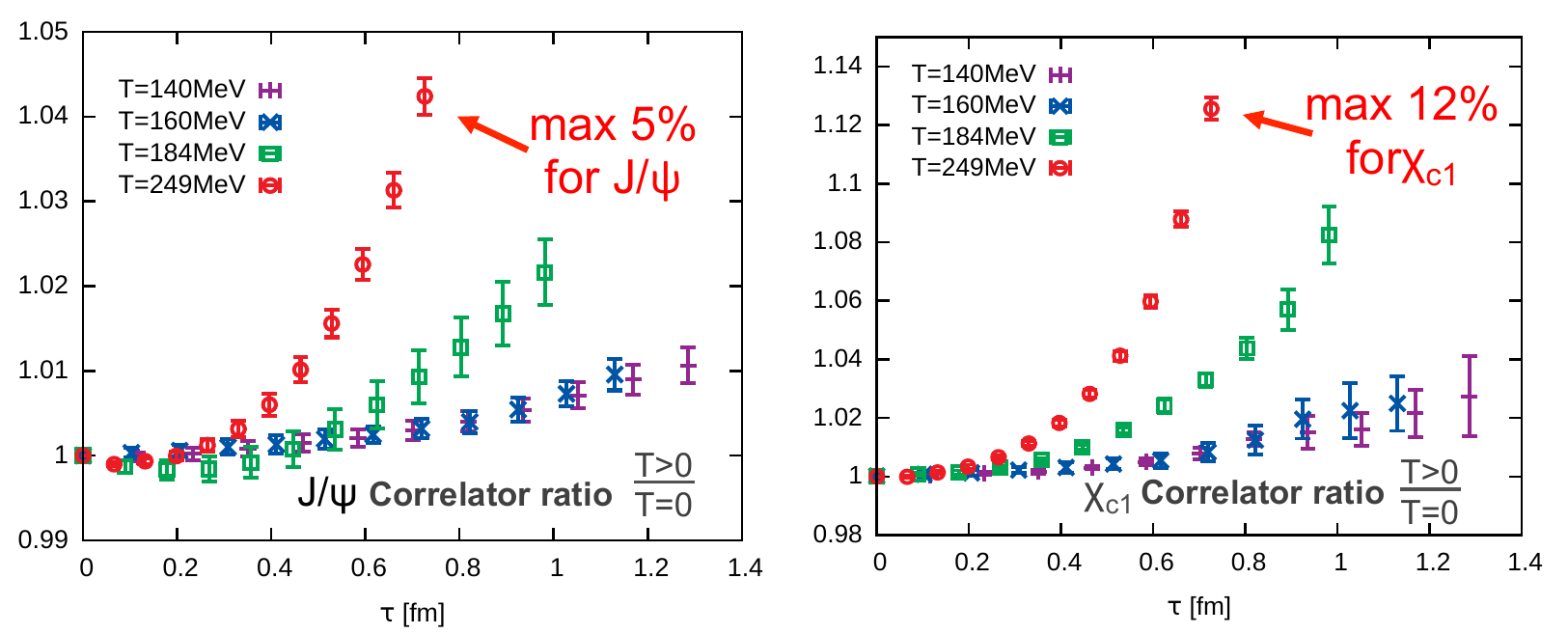}
  \end{center}\vspace{-0.7cm}
 \caption{Ratio of $T>0$ and $T\approx0$ heavy $Q\bar{Q}$ correlation functions in Euclidean time. The strength of in-medium modification is sequentially ordered according to the vacuum binding energy. }\label{Fig:CompFiniteVac}\vspace{-0.3cm}
\end{wrapfigure}

Here we focus on two recent results obtained from an ongoing lattice NRQCD study on in-medium quarkonium. First Bottomonium results have been published in \cite{Kim:2014iga, Kim:2014fta} and preliminary results on Charmonium in \cite{Kim:2015csj}. In Fig.\ref{Fig:CompFiniteVac} we show the ratio of in-medium $Q\bar{Q}$ correlators to their $T=0$ values in different

\setlength{\columnsep}{12.5pt}
\noindent channels. If the ratio is unity, no in-medium modification is present. We find a common non-monotonic behavior among all $Q\bar{Q}$ species. Just below the transition temperature the ratio lies slightly above one, it moves closer to unity at $T\to T_{\rm pc}$ and then begins to bend upwards as temperature is further increased. The absolute effect we observe is quite small, at the highest temperature $T=249$MeV it is up to $1\%$ for $\Upsilon(1S)$, $5\%$ for $\chi_b(1P)$ and $J/\psi$ and $12\%$ for $\chi_c(1P)$. Our main finding is that as expected for sequential suppression, 
the size of in-medium modification of the correlator ratio follows the hierarchy of the 
$T=0$ binding energy of the corresponding bound states. For the states investigated here this hierarchy reads $E^{\rm \Upsilon(1S)}_{\rm b}\simeq 1.1{\rm GeV} >  E^{\rm \chi_b(1P)}_{\rm b} \approx E^{\rm J/\psi(1S)}_{\rm b} \simeq 640{\rm MeV} > E^{\rm \chi_c(1P)}_{\rm b} \simeq 200{\rm MeV}$.

\setlength{\columnsep}{14pt}
\begin{wrapfigure}{r}{0.52\textwidth}
\vspace{-0.9cm}  \begin{center}
\hspace{-0.4cm}\includegraphics[scale=0.45, trim=0.2cm 0 0 0, clip=true]{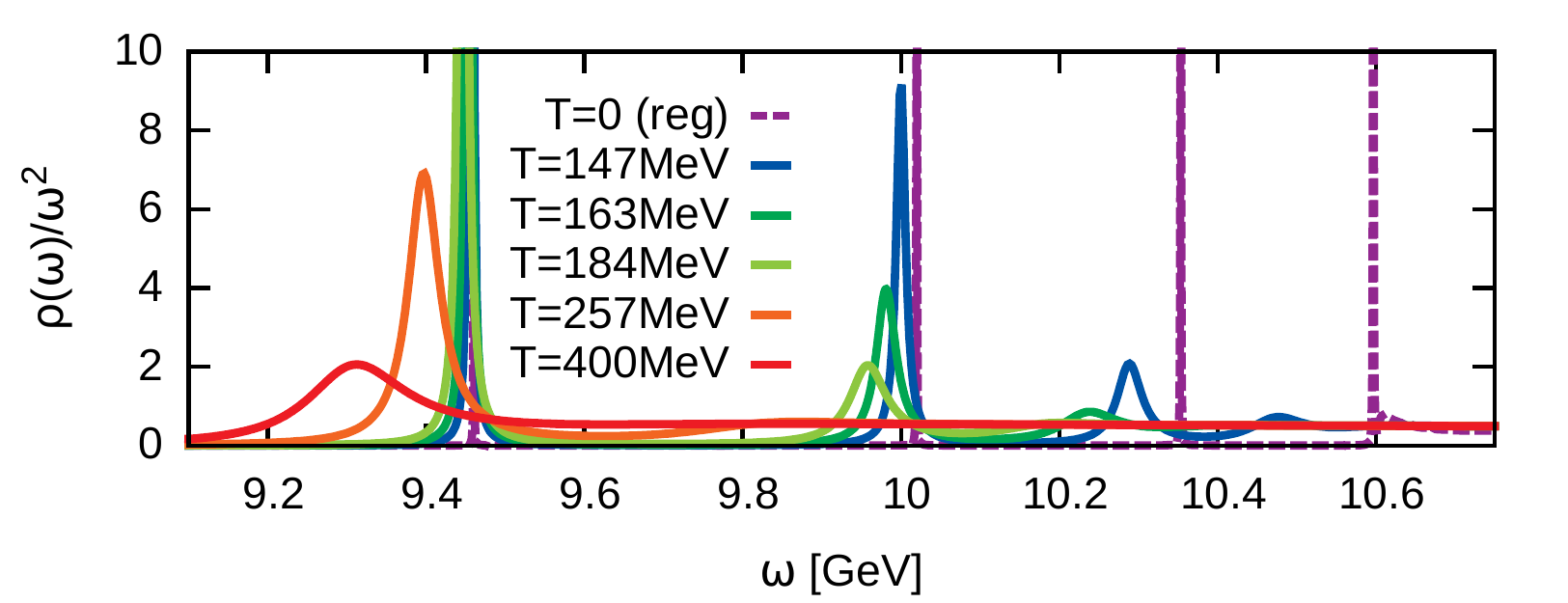}\vspace{-0.6cm}
  \includegraphics[scale=0.59]{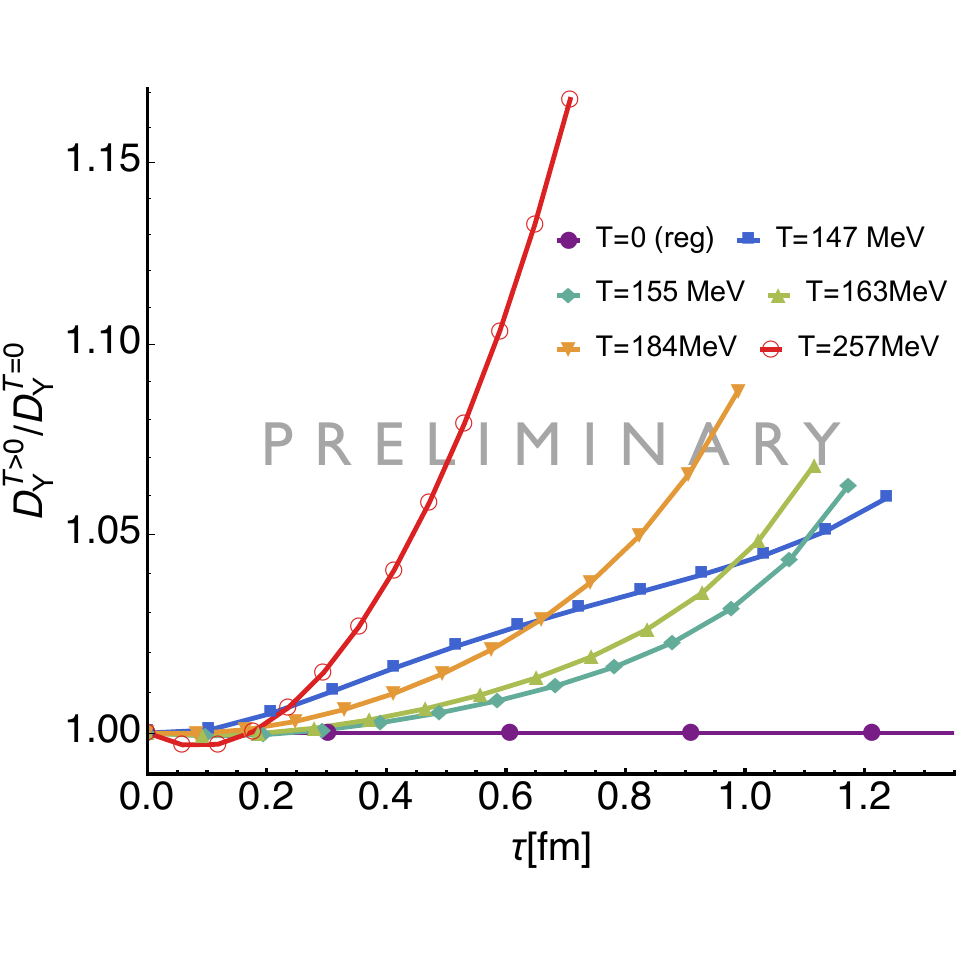}
  \end{center}\vspace{-1.3cm}
 \caption{(top) $\Upsilon(1S)$ spectral functions from a state-of-the-art lattice QCD potential based study \cite{Burnier:2015tda}. (bottom) A preliminary computation of the corresponding correlator ratios, which show the same non-monotonic behavior as those in our lattice NRQCD study.}\label{Fig:CompPotComp}\vspace{-0.5cm}
\end{wrapfigure}
To interpret the physics encoded in these ratios, we take a look at ratios that follow from $b\bar{b}$ S-wave spectra obtained via a state-of-the-art lattice QCD potential based study \cite{Burnier:2015tda}. As shown in the top panel of Fig.\ref{Fig:CompPotComp}, the potential based spectra show a clear sequential melting pattern with a shift of $m_{Q\bar{Q}}$ to lower values. If we compute the corresponding correlator ratios from appropriately regularized Laplace transforms, similar non-monotonic behavior around $T_{\rm pc}$ seems to emerge. This outcome supports the interpretation of the lattice NRQCD ratios as manifestations of sequential state melting.

Let us now inspect the spectra obtained from a Bayesian reconstruction of NRQCD in-medium correlators. It is known that Bayesian reconstructions suffer both from a limited number of datapoints, as well as from the finite physical extent available on the lattice at $T>0$.  As shown in Refs.~\cite{Kim:2014iga, Kim:2014fta} we have assessed these effects by taking $T=0$ correlator sets, truncating them to the number of points available at finite temperature and then feeding the truncated data to the novel Bayesian reconstruction prescription. In turn we set stringent accuracy limits for our approach to detect in-medium mass shifts and width broadening. It turns out that the minute changes we observe in our reconstructed ground state peaks at $T>0$ are too small at our current statistics to be unambiguously attributed to in-medium effects. Therefore we limit ourselves here to the question of state survival.

 To answer whether a bound state signal is present in a reconstructed spectrum we have to recognize 
  
\setlength{\columnsep}{9pt}
\begin{wrapfigure}{r}{0.41\textwidth}
 \begin{center}\vspace{-1.1cm}
\includegraphics[scale=0.77, trim=0.8cm 0 0 0, clip=true]{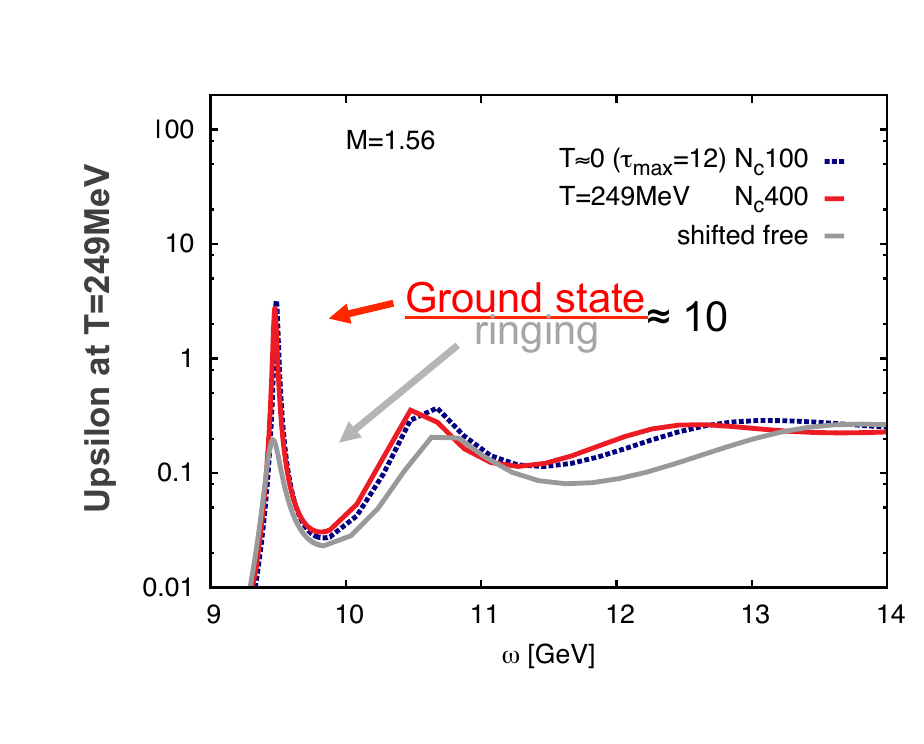}\vspace{-0.7cm}
  \includegraphics[scale=0.77, trim=0.1cm 0 0 0, clip=true]{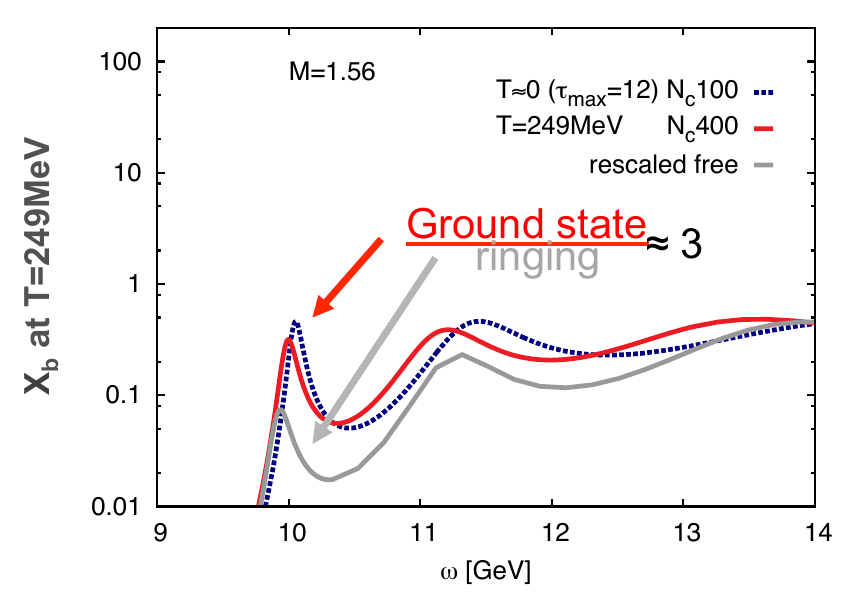}\vspace{-0.1cm}
  \includegraphics[scale=0.82, trim=0.2cm 0 0 0, clip=true]{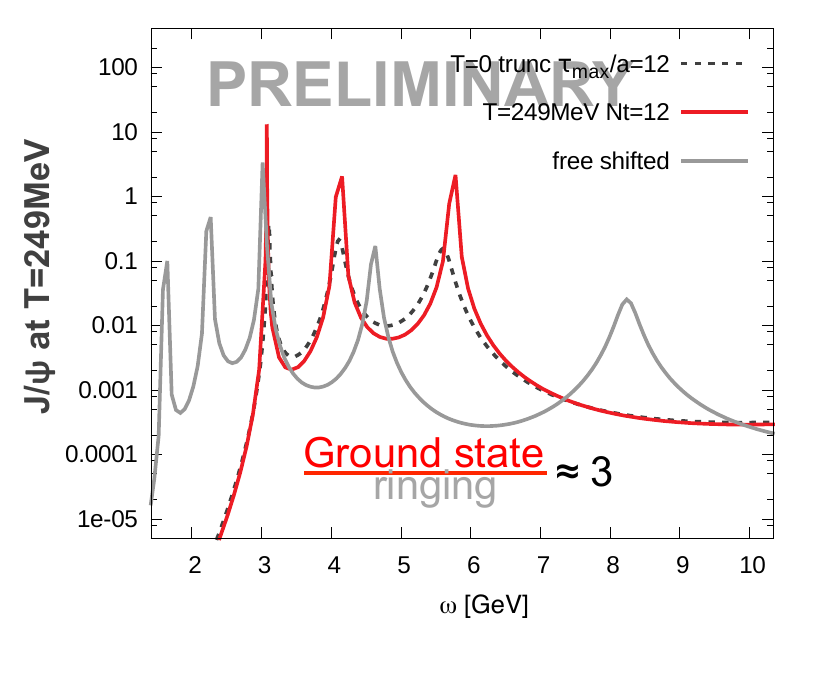}
  \end{center}\vspace{-1.2cm}
 \caption{Reconstruction of fully interacting (red) and free NRQCD spectra (gray), as well as from the truncated $T=0$ correlator (blue, dashed). (top) $\Upsilon(1S)$ spectrum (center) $\chi_b(1P)$ spectrum (bottom) preliminary $J/\psi (1S)$ spectrum.  }\label{Fig:SpecFreeInt}\vspace{-1.7cm}
\end{wrapfigure}\noindent artifacts that could mimic such a signal. One prominent issue is that of numerical ringing, well known as the Gibbs phenomenon in the closely related inverse Fourier transform. We propose to identify the physical signal by comparing the interacting spectra to the reconstruction of non-interacting spectra. The corresponding free correlator can be measured by setting all links on the underlying lattice to unit matrices. It can be analytically computed and its spectrum is known to be devoid of bound state peaks. Due to the finite number of points available however its Bayesian reconstruction will contain wiggly features that have to be identified as numerical artifacts. 
	
In Fig. \ref{Fig:SpecFreeInt} we compare at $T=249$MeV the reconstructions of interacting (red) and appropriately shifted non-interacting spectra (gray). In addition the reconstruction from the truncated $T=0$ correlator is given in blue (dashed). Note that the blue curves deviate only slightly from the actual $T=0$ spectra and so do also the red curves \cite{Kim:2014iga, Kim:2014fta}. We find that the free spectra do contain unphysical peaked structures at low frequencies. A direct comparison for $\Upsilon(1S)$ shows that the bound state signal is still an order of magnitude stronger than ringing and the state clearly survives. For $\chi_b(1P)$ as expected, the difference is much smaller but still a factor three remains, which we interpret as survival of the bound state. In particuar the difference between the $T>0$ (red) and truncated $T=0$ result (blue) is again minimal. Our preliminary finding for $J/\psi$ shows also roughly a factor of three between numerical ringing and the bound state signal, which hints at a survival of this state deep into the QGP phase.

It will be the task of future work to quantitatively determine the in-medium properties of these surviving states, i.e. both their in-medium mass shifts and width broadening. To this end we are in the process of significantly increasing the statistics for the NRQCD correlators.

SK is supported by Korean NRF grant No.\  2015R1A2\\ A2A01005916,  PP by the U.S. DOE under contract DE-\\ SC0012704. We thank HotQCD for providing the gauge\\ configurations for this study.

\end{document}